\documentclass[a4paper,aps,prd,onecolumn,preprintnumbers,showpacs,superscriptaddress,nofootinbib]{revtex4}

\usepackage{graphics,graphicx,epsfig}
\usepackage{amsfonts,amsmath,amssymb}

\begin{document}

\title{Shadow of a rotating braneworld black hole}
\author{Leonardo Amarilla}
\email{yellow@df.uba.ar} \affiliation{Departamento de F\'{\i}sica, Facultad de Ciencias Exactas y Naturales, Universidad de Buenos Aires, Ciudad Universitaria, Pabell\'on 1, 1428, Buenos Aires, Argentina.}
\author{Ernesto F. Eiroa} \email{eiroa@iafe.uba.ar} \affiliation{Instituto de Astronom\'{\i}a y F\'{\i}sica del Espacio, C.C. 67 Suc. 28, 1428, Buenos Aires, Argentina.} \affiliation{Departamento de F\'{\i}sica, Facultad de Ciencias Exactas y Naturales, Universidad de Buenos Aires, Ciudad Universitaria, Pabell\'on 1, 1428, Buenos Aires, Argentina.}


\begin{abstract}
We investigate the shadow cast by a rotating braneworld black hole, in the Randall-Sundrum scenario. In addition to the angular momentum, the tidal charge term deforms the shape of the shadow. For a given value of the rotation parameter, the presence of a negative tidal charge enlarges the shadow and reduces its deformation with respect to Kerr spacetime, while for a positive charge, the opposite effect is obtained. We also analyze the case in which the combination of the rotation parameter and the tidal charge results in a naked singularity. We discuss the observational prospects corresponding to the supermassive black hole at the Galactic center.
\end{abstract}

\maketitle

\section{Introduction}

In braneworld cosmologies, the ordinary matter is on a three dimensional space (the brane) which is embedded in a larger space (the bulk) where only gravity can propagate. These models, which have received great attention in recent years \cite{bwrev}, were proposed to solve the hierarchy problem (the difficulty in explaining why the gravity scale is sixteen orders of magnitude greater than the electro-weak scale) and they have motivation in recent developments of string theory, known as M-theory. The presence of the extra dimensions would modify the properties of black holes \cite{kanti}. Clancy et al. \cite{cgl} showed that, in the simplest of braneworld scenarios, the Randall-Sundrum \cite{rsII} models (a positive tension brane in a bulk with one extra dimension and a negative cosmological constant) primordial black holes formed in the high energy epoch would have a longer lifetime, because of a different evaporation law. Braneworld primordial black holes could have a growth of their mass through accretion of surrounding radiation during the high energy phase, increasing their lifetime \cite{majumdar}. These black holes could have survived up to the present and they have an induced four dimensional metric on the brane different from the Schwarzschild metric. In these models,  black holes also may be created in high energy collisions in particle accelerators or in cosmic rays \cite{kanti}. In the Randall-Sundrum scenario, Dadhich et al.  found a spherically symmetric black hole solution on a three dimensional brane \cite{dmpr}, which is characterized by a tidal charge, due to gravitational effects coming from the fifth dimension. Recently, rotating black hole solutions with a tidal charge were studied by Aliev et al. \cite{aliev1,aliev2}.

The study of the null geodesics around black holes, in which photons coming from astrophysical sources move, is useful to obtain information about these objects. Gravitational lensing by black holes has been investigated by several authors in the last decade, mainly because of the strong evidence about the presence of supermassive black holes at the center of galaxies. A useful approximate analytical method for obtaining the positions, magnifications, and time delays of the relativistic images corresponding to black hole lenses, is the strong deflection limit. It was introduced by Darwin \cite{darwin} for the Schwarzschild geometry, rediscovered several times \cite{otros}, extended to the Reissner-Nordstr\"om spacetime \cite{eiroto}, and to any spherically symmetric black holes \cite{boz}. Numerical studies \cite{numerical} of black hole lenses were performed too. Non-rotating braneworld black holes were studied as gravitational lenses \cite{bwlens} as well. Kerr black hole lenses were considered by several authors \cite{bozza1,bozza2,vazquez,kraniotis}.  Rotating black holes present apparent shapes or shadows with an optical deformation due to the spin \cite{bardeen,chandra}, instead of being circles as in the case of non-rotating ones. This topic has been reexamined by several authors in the last few years \cite{falcke,devries,takahashi,bozza2,hioki,bambi,maeda,amarilla,zakharov}, with the expectation that the direct observation of black holes will be possible in the near future \cite{zakharov}; therefore the study of the shadows will be useful for measuring the properties of astrophysical black holes. Optical properties of rotating braneworld black holes were studied by Schee and Stuchlik \cite{schee}. For more details about black hole gravitational lensing and a discussion of its observational prospects, see the recent review article \cite{bozzareview} and the references therein.

In this paper, we study how the presence of the tidal charge modifies the form of the shadow cast by the spinning black hole with a tidal charge introduced by Aliev et al. \cite{aliev1,aliev2}, in the Randall-Sundrum braneworld scenario. This topic has been previously considered by Schee and Stuchlik \cite{schee}; in our work we calculate the observables defined by Hioki and Maeda \cite{maeda} and we concentrate on the analysis of the shadow corresponding to the Galactic black hole. When the appropriate combination of the rotation parameter and the tidal charge is large enough, the event horizon disappears and a naked singularity is obtained. This case, not analyzed in Ref. \cite{schee}, is considered here. In Sec. 2, we review the basic aspects of the geometry and the geodesics of the rotating braneworld black hole. In Sec. 3, we obtain the shadows of black holes with different values of the rotation parameter and the tidal charge. In Sec. 4, the study is extended to naked singularities. Finally, in Sec. 5, we discuss the results found. We adopt units such that $G=c=1$.

\section{Spinning braneworld black holes}

We adopt for the rotating black hole in the Randall-Sundrum braneworld scenario the metric introduced in Ref. \cite{aliev1}, which in the Boyer-Lindquist coordinates has the form
\begin{equation}
ds^2  =  -\frac{\Delta}{\Sigma} \left(dt - a \sin^2\theta d\phi \right)^2 + \Sigma \left(\frac{dr^2}{\Delta} + d\theta^{2}\right) + \frac{\sin^2\theta}{\Sigma}\left[ a dt - (r^2+a^2) d\phi \right]^2 ,
\label{metric}
\end{equation}
with
\begin{equation}
\Delta = r^2 + a^2 -2 M r+ Q , \;\;\;\;\; \Sigma=r^2+a^2 \cos^2\theta ,
\nonumber
\label{aux}
\end{equation}
where $ M $ is  the mass, $ a $ is the rotation parameter (angular momentum per unit mass, $ a=J/M $), and $ Q $ is the tidal charge of the black hole. This metric was obtained by adopting the assumption that the induced geometry on the 3D brane has the Kerr-Schild form \cite{aliev1}. It is unclear if this choice of the metric on the brane is indeed fulfilled by an exact bulk metric  \cite{aliev1}. The tidal charge is interpreted as an imprint of the gravitational effects from the bulk space \cite{aliev1}. We let the tidal charge take any sign, but some authors argue that the negative sign is more natural (see \cite{aliev1,aliev2} and references therein). When $Q=0$ one recovers the Kerr geometry. The non-rotating case, i.e. $a=0$, corresponds to the brane black hole metric previously studied in Ref. \cite{dmpr}. The properties of the geometry (\ref{metric}) are similar to those of the Kerr-Newman metric in general relativity. The event horizon is determined by largest root of the equation $\Delta=0$, given by  
\begin{equation}
r_{+}= M + \sqrt{M^2 - a^2- Q}.
\label{horizon}
\end{equation}
Thus, the event horizon exists if $Q  \leq Q_c =M^2 -a^2$, with the equal sign corresponding to a maximally rotating black hole (with $r_{+} = M $). When the tidal charge is positive, this condition leads to the Kerr type bound on the angular momentum, i.e. the rotation parameter cannot exceed the mass. But the situation is different for the negative tidal charge, for which the maximally rotating black hole has the rotation parameter greater than the mass. In what follows, for simplicity, we adimensionalize all quantities with the mass of the black hole, which is equivalent to put $M=1$ in all equations.\\

When a black hole is between a source of light and an observer, the light reaches the observer after being deflected by the black hole gravitational field; but some part of the photons emitted by the source, those with small impact parameters, end up falling into the black hole, not reaching the observer, giving as result a dark zone in the sky called the shadow. The apparent shape of a black hole is thus defined by the boundary of the shadow. In order to obtain the apparent shape, we need to study the geodesic structure. The Hamilton-Jacobi equation determines the geodesics for a given geometry:
\begin{equation}  \label{SHJ1}
\frac{\partial S}{\partial \lambda}=-\frac{1}{2}g^{\mu\nu}\frac{\partial S}{\partial x^{\mu}}\frac{\partial S}{\partial x^{\nu}},
\end{equation}
where $\lambda$ is an affine parameter along the geodesics, $g_{\mu\nu}$ are the components of the metric tensor and $S$ is the Jacobi action. When the problem is \textit{separable}, the Jacobi action $S$ can be written in the form
\begin{equation}  \label{SHJ2}
S=\frac{1}{2}m^2 \lambda - E t + L_z \phi + S_{r}(r)+S_{\theta}(\theta).
\end{equation}
where $m$ is the mass of a test particle. The second term on the right hand side is related to the conservation of energy $E$, while the third term is related to the conservation of the angular momentum in the direction of the axis of symmetry $L_z$. In the case  of null geodesics, we have that $m =0$, and from the Hamilton-Jacobi equation, the following equations of motion are obtained \cite{aliev2}:
\begin{equation}
\Sigma\frac{dt}{d\lambda}=a(L_z -aE\sin^2\theta)+
\frac{r^2+a^2}{\Delta}\left[(r^2+a^2)E -a L_z \right],
\label{et}
\end{equation}
\begin{equation}
\Sigma\frac{d\phi}{d\lambda}=\left(\frac{L_z}{\sin^2\theta}
-aE\right)+\frac{a}{\Delta}\left[(r^2+a^2)E -a L_z \right],
\label{ephi}
\end{equation}
\begin{equation}
\Sigma\frac{dr}{d\lambda}=\sqrt{\mathcal{R}},
\label{er}
\end{equation}
\begin{equation}
\Sigma\frac{d\theta}{d\lambda}=\sqrt{\Theta},
\label{etheta}
\end{equation}
where the functions $\mathcal{R}(r)$ and $\Theta(\theta)$ are defined by
\begin{equation}
\mathcal{R}=\left[(r^2+a^2)E -a L_z \right]^2-\Delta\left[\mathcal{K}+(L_z -a E)^2\right],
\end{equation}
and
\begin{equation}
\Theta=\mathcal{K}+\cos^2\theta\left(a^2E^2-\frac{L_z^2}{\sin^2\theta}\right),
\end{equation}
with $ \mathcal{K} $ a constant of separation. These equations determine the propagation of light in the spacetime of the braneworld rotating black hole. The geometry (\ref{metric}) is asymptotically flat, so the trajectory of photons are straight lines at infinity. The light rays are, in general, characterized by two impact parameters, which can be expressed in terms of the constants of motion $E$, $L_z$ and the Carter constant $\mathcal{K}$. Combining these quantities we define as usual  $\xi=L_z/E$ and $\eta=\mathcal{K}/E^{2}$, which are the impact parameters for general orbits around the black hole. We use Eq. (\ref{er}) to derive the orbits with constant $r$ in order to obtain the boundary of the shadow of the black hole. These orbits satisfy the conditions $\mathcal{R}(r)=0=d\mathcal{R}(r)/dr$, which are fulfilled by the values of the impact parameters that determine the contour of the shadow, namely, 
\begin{equation}
\xi(r)=\frac{a^{2}(1+r)+r\left[ (r-3)r+2 Q \right] }{a(1-r)}=\xi_\text{K}(r)-\frac{2 Q r}{a(r-1)},
\label{eqxi}
\end{equation}
\begin{equation}
\eta(r)=\frac{r^{2}\left\{ 4a^{2}(r-Q)-\left[ r(r-3)+2 Q \right] ^{2}\right\} }{a^{2}(r-1)^{2}}=\eta_\text{K}(r)-\frac{4 Q r^{2}(\Delta-r)}{a^{2}(r-1)^{2}},
\label{eqeta}
\end{equation}
where
\begin{equation*}
\xi_{\text{K}}(r)=\frac{r^{2}-r \Delta _{\text{K}}-a^{2}}{a(r-1)},
\end{equation*}
\begin{equation*}
\eta_{\text{K}}(r)=\frac{r^{3}\left[ 4 \Delta _{\text{K}} -r(r-1)^{2}\right] }{a^{2}(r-1)^{2}}
\end{equation*}
(with $\Delta _{\text{K}} = \Delta |_{Q=0}$) are the corresponding values in Kerr geometry.

\section{Black hole shadow}

To describe the shadow, we adopt the celestial coordinates (see for example \cite{vazquez}):
\begin{equation}  \label{alpha}
\alpha=\lim_{r_{0}\rightarrow \infty}\left( -r_{0}^{2}\sin\theta_{0}\frac{d\phi}{dr}\right)
\end{equation}
and
\begin{equation}  \label{beta1}
\beta=\lim_{r_{0}\rightarrow \infty}r_{0}^{2}\frac{d\theta}{dr},
\end{equation}
where $r_{0}$ goes to infinity because we consider an observer far away from the black hole, and $\theta_{0}$ is the angular coordinate of the observer, i.e. the inclination angle between the rotation axis of the black hole and the line of sight of the observer. The coordinates $\alpha $ and $\beta$ are the apparent perpendicular distances of the image as seen from the axis of symmetry and from its projection on the equatorial plane, respectively. If we calculate $d\phi/dr$ and $d\theta/dr$ from the metric given by Eq. (\ref{metric}) and take the limit of a far away observer, we have that, as a function of the constants of motion, the celestial coordinates take the form
\begin{equation}  \label{alphapsi1}
\alpha=-\xi\csc\theta_{0}
\end{equation}
and
\begin{equation}  \label{beta2}
\beta=\pm \sqrt{\eta + a^{2}\cos ^{2}\theta_{0}-\xi^{2}\cot ^{2}
\theta_{0}},
\end{equation}
where Eqs. (\ref{ephi}), (\ref{er}), and (\ref{etheta}) were used to calculate $d\theta /dr$ and $d\phi /dr$. These equations have implicitly the same form as for the Kerr metric,  with the new $\xi $ and $\eta$ given by Eqs. (\ref{eqxi}) and (\ref{eqeta}) (a detailed calculation of the values of $\xi$ and $\eta$, and the expressions of the celestial coordinates $\alpha$ and $\beta$ as a function of the constants of motion for Kerr geometry, are given in \cite{vazquez}).

For the characterization of the form of the shadow, we adopt the observables defined in \cite{maeda}: the radius $R_{s}$ and the distortion parameter $\delta _{s}$. The observable $R_s$ is the radius of a reference circle passing by three points: the top position $(\alpha_t, \beta_t)$ of the shadow, the bottom position $(\alpha _b,\beta _b)$ of the shadow, and the point corresponding to the unstable retrograde circular orbit seen from an observer on the equatorial plane $(\alpha _r,0)$. The distortion parameter is defined by $D/R_s$, where $D$ is the difference between the
endpoints of the circle and of the shadow, both of them at the opposite side of the point $(\alpha _r,0)$, i.e. corresponding to the prograde circular orbit. The radius $R_s$ basically gives the approximate size of the shadow, while $\delta_s$ measures its deformation with respect to the reference circle (see \cite{maeda} for more details). If the inclination angle $\theta _{0}$ is independently known (see for example \cite{li-narayan}), precise enough measurements of $R_{s}$ and $\delta _{s}$ could serve to obtain the rotation parameter $a$ and the tidal charge $Q$ (both adimensionalized with the black hole mass, as stated above). A simple way to extract this information is by plotting the contour curves of constant $R_s$ and $\delta _s$ in the plane $(a,Q)$; the point in the plane where they intersect gives the corresponding values of the rotation parameter $a$ and the tidal charge $Q$.

\begin{figure}[t!]
\begin{center}
\includegraphics[width=0.4\linewidth]{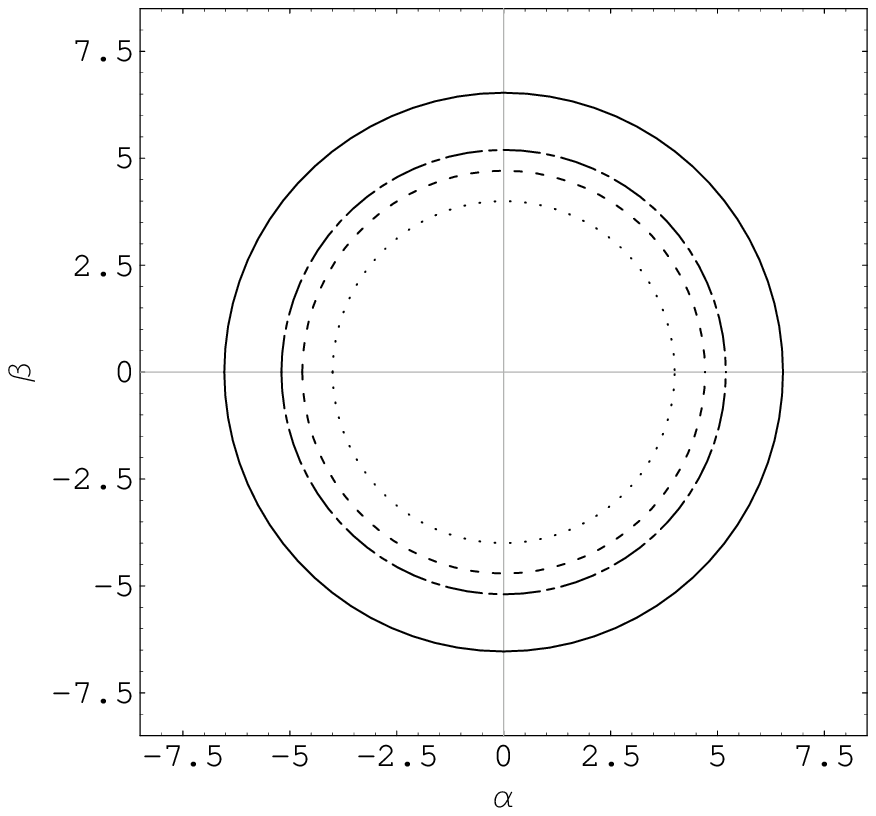}
\hspace{0.05\linewidth}
\includegraphics[width=0.4\linewidth]{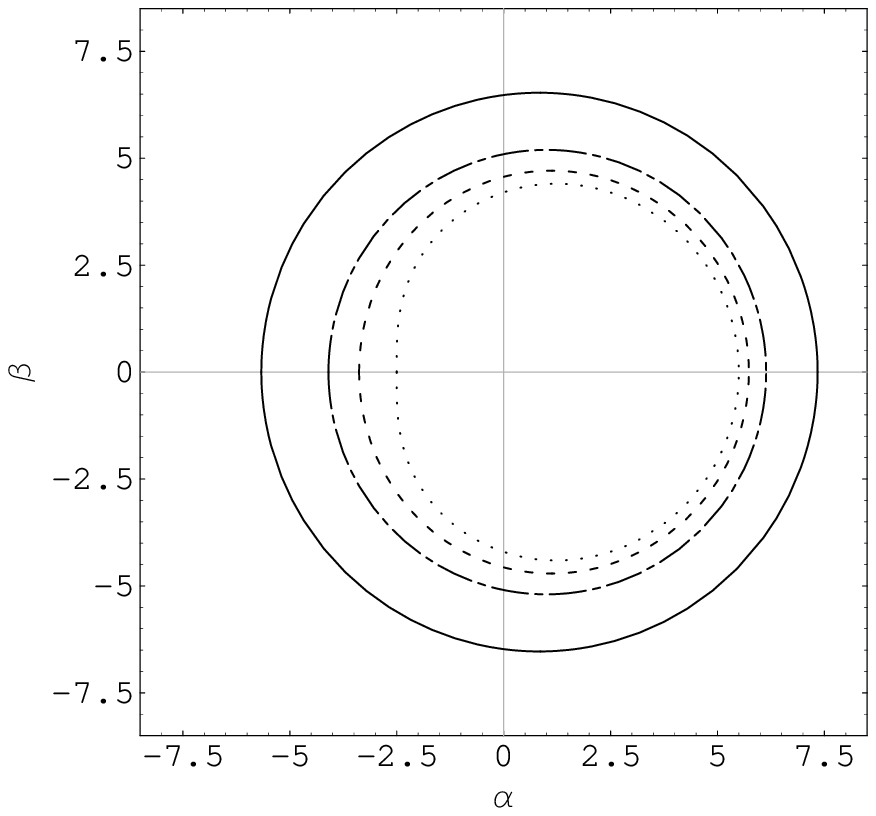}\\
\includegraphics[width=0.4\linewidth]{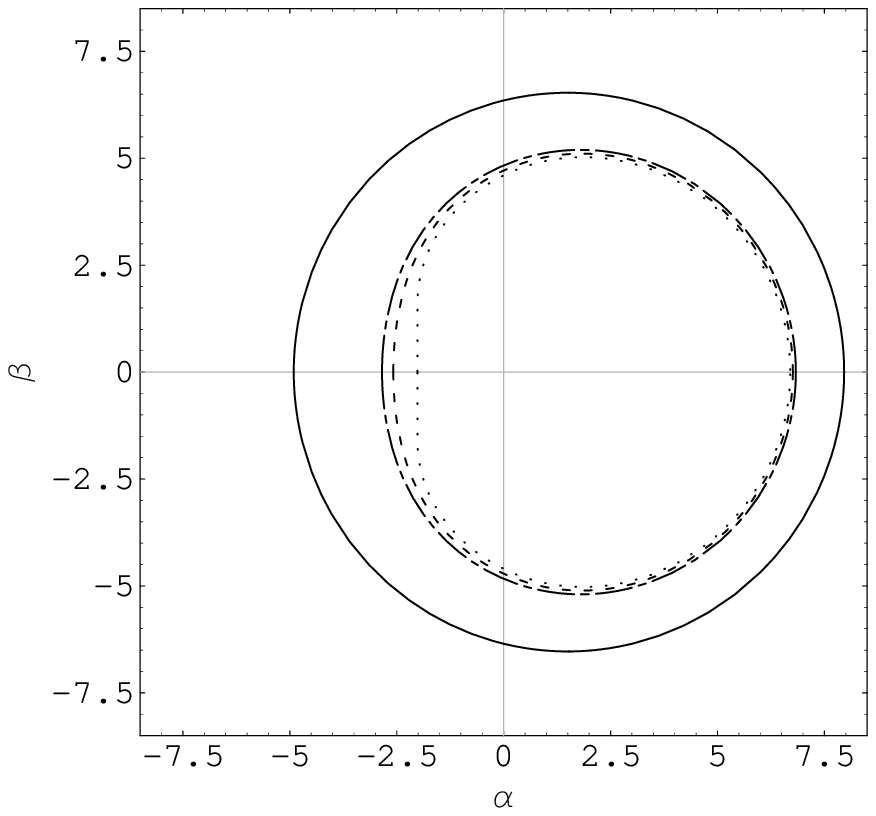}
\hspace{0.05\linewidth}
\includegraphics[width=0.4\linewidth]{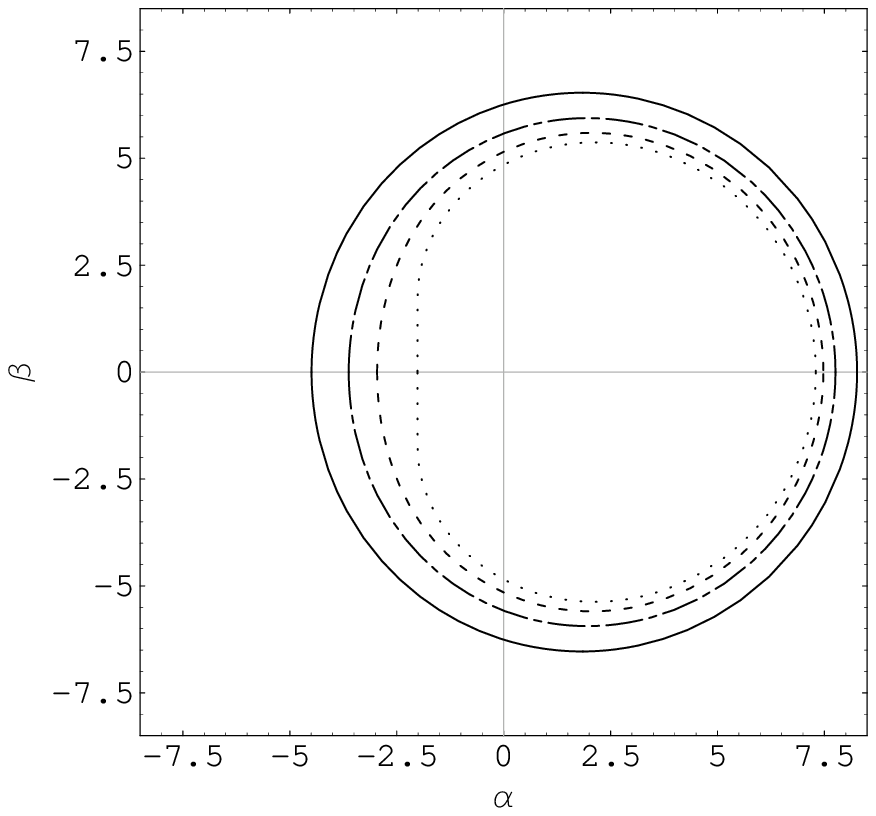}
\end{center}
\caption{Silhouette of the shadow cast by a black hole situated at the origin of coordinates with inclination angle $\theta _0=\pi /2$, having a rotation parameter $a$ and a tidal charge $Q$. Upper row, left: $a=0$, $Q= -2$ (full line), $0$ (dashed-dotted line), $0.5$ (dashed line), and $Q_{c}=1$ (dotted line). Upper row, right: $a=0.5$, $Q= -2$ (full line), $0$ (dashed-dotted line), $0.5$ (dashed line), and $Q_{c}=0.75$ (dotted line). Lower row, left:  $a=0.9$,   $Q= -2$ (full line), $0$ (dashed-dotted line), $0.1$ (dashed line), and $Q_{c}=0.19$ (dotted line). Lower row, right:  $a=1.1$,   $Q= -2$ (full line), $-1$ (dashed-dotted line), $-0.5$ (dashed line), and $Q_{c}=-0.21$  (dotted line). The shadow corresponds to each curve and the region inside it. All quantities were adimensionalized with the mass of the black hole (see text).}
\label{fig1}
\end{figure}

When the observer is situated in the equatorial plane of the black hole, the inclination angle is $\theta_{0}=\pi/2$ and the gravitational effects on the shadow, which grow with $\theta _0$, are larger. The inclination angle corresponding to the Galactic supermassive black hole is also expected to lie close to $\pi /2$. In this interesting case, we have simply
\begin{equation}  \label{alphapsi2}
\alpha=-\xi
\end{equation}
and
\begin{equation}  \label{beta3}
\beta=\pm \sqrt{\eta}.
\end{equation}
For the visualization of the shape of the black hole shadow one needs to plot $\beta$ vs $\alpha $. In Fig. \ref{fig1}, we show the contour of the shadows of black holes with rotation parameters $a=0$ (upper row, left), $a=0.5$ (upper row, right), $a=0.9$ (lower row, left), and $a=1.1$ (lower row, right), for several values of the tidal charge $Q$. The effect of a positive $Q$ is to decrease the size of the shadow, as in the Kerr-Newman case for the square of the electromagnetic charge \cite{devries}, while negative values generate an enlargement of its size. The plots of the contours of the shadows displayed in Fig. \ref{fig1} are similar to those previously obtained in Ref. \cite{schee} for other values of the parameters.

\begin{figure}[t!]
\begin{center}
\includegraphics[width=0.4\linewidth]{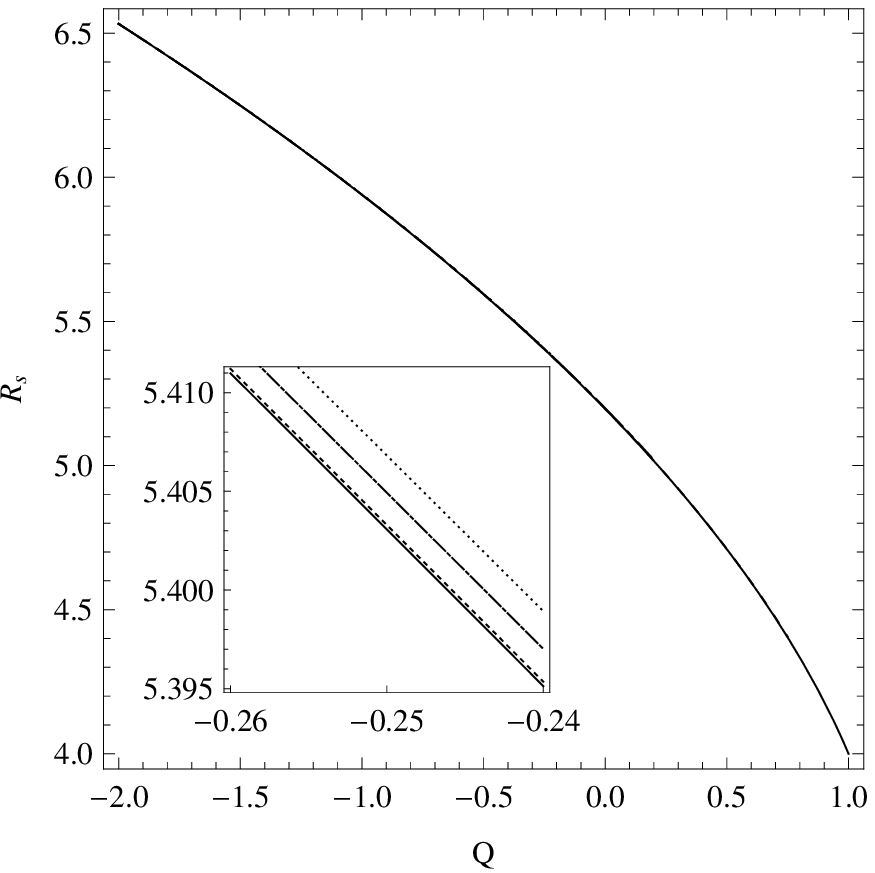}
\hspace{0.05\linewidth}
\includegraphics[width=0.4\linewidth]{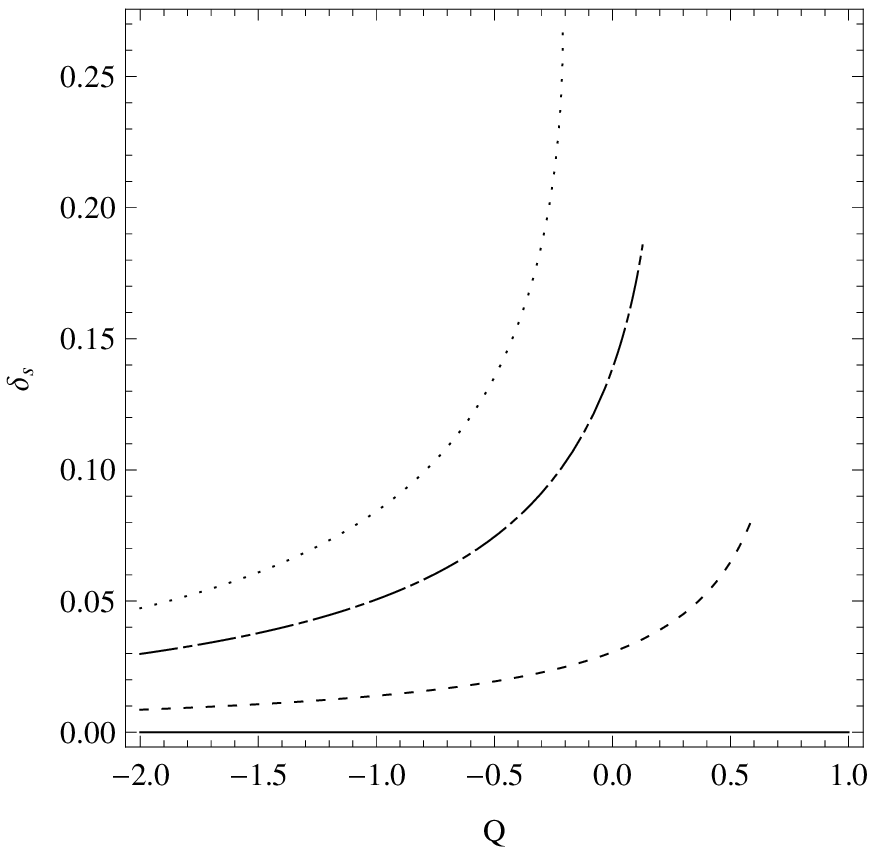}
\end{center}
\caption{Observables $R_{s}$ and $\delta _{s}$ as functions of the tidal charge $Q $, corresponding to the shadow of a black hole situated at the origin of coordinates with inclination angle $\theta _0=\pi /2$, and spin parameters $a=0$ (full line), $a=0.5$ (dashed line), $a=0.9$ (dashed-dotted line), and $a=1.1$ (dotted line). The curves with different values of $a$ are not distinguishable in the left plot;  a smaller range of $Q$ is shown in the frame inside, in which the different curves can be appreciated. }
\label{fig2}
\end{figure}

\begin{figure}[t!]
\begin{center}
\includegraphics[width=0.4\linewidth]{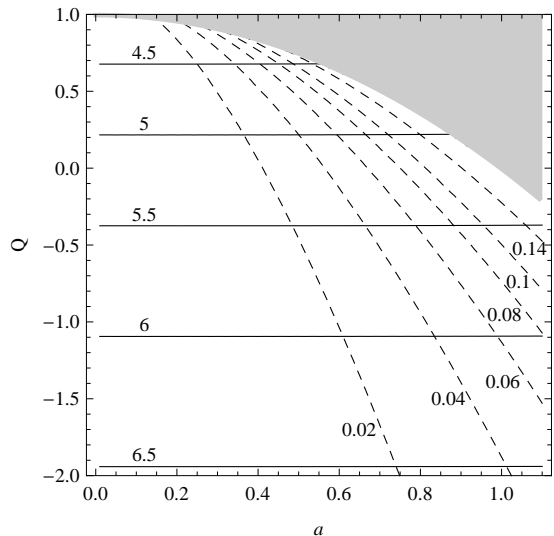}
\end{center}
\caption{Contour plots of the observables $R_{s}$ and $\delta _{s}$ in the plane $(a,Q)$, for a black hole situated at the origin of coordinates with inclination angle $\theta _0=\pi /2$. Each curve is labeled with the corresponding value of $R_{s}$ or $\delta _{s}$. The light gray zone represents naked singularities.}
\label{fig3}
\end{figure}

The observable $R_s$ can be calculated from the equation
\begin{equation*}
R_{s}=\frac{(\alpha _t -\alpha_r)^2 + \beta_t ^2}{2|\alpha _t -\alpha_r|},
\end{equation*}
and the observable $\delta_s$ is given by
\begin{equation*}
\delta _s=\frac{\tilde{\alpha}_p - \alpha_p}{R_{s}},
\end{equation*}
where $(\tilde{\alpha}_p, 0)$ and $(\alpha_p, 0)$ are the points where the reference circle and the contour of the shadow cut the horizontal axis at the opposite side of $(\alpha_r, 0)$, respectively. In Fig. \ref{fig2},  the observables $R_{s}$ and $\delta_{s}$ are shown as functions of the tidal charge $Q$, for several values of the rotation parameter of the black hole: $a=0$ (full line), $a=0.5$ (dashed line), $a=0.9$ (dashed-dotted line), and $a=1.1$ (dotted line). From Fig. \ref{fig2}, we see that for different values of $a$, the observable $R_{s}$ has a similar behavior as a function of $Q$ and the curves are not distinguishable. In the frame inside, the range in $Q$ is smaller, and the curves corresponding to different rotation parameters can be appreciated; it can be seen that, for a fixed value of $Q$, the difference between the $a=0$ curve and the $a=1.1$ one is of order $10^{-3}$, leading to a small variation in the size of the shadow as a function of $a$, as expected from Kerr and Kerr-Newman case. As previously stated, positive values of $Q$ generate a decrease in the size (and then in $R_{s}$) of the shadow, and negative values generate an enlargement of it. In Fig. \ref{fig2}, the distortion parameter $\delta_{s}$ keeps null for $a=0$ (full line), as expected for the non rotating case, and increases as a function of $Q$ for $a=0.5$ (dashed line), $a=0.9$ (dashed-dotted line),  and $a=1.1$ (dotted line). The distortion is maximal when $Q$ reaches its limiting value $Q_{c}$. When the tidal charge takes negative values, the distortion in the shape of the shadow with respect to its reference circle decreases as $Q$ gets more negative. For fixed $Q$, the deformation of the shadow increases with $a$. The contour curves of constant $R_s$ and $\delta _s$ in the plane $(a,Q)$ are shown for some representative values in Fig. \ref{fig3}. As mentioned above, if $R_s$ and $\delta  _s$ are obtained from observations, the point in the plane where the associated contour curves intersect gives the corresponding values of the rotation parameter $a$ and the tidal charge $Q$. 

Summarizing, for a fixed value of $a$, the presence of a negative (positive) tidal charge leads to a larger (smaller) shadow than in the case of Kerr geometry, corresponding to a larger (smaller) value of $R_{s}$;  while a negative (positive) value of $Q$ gives a less (more) distorted shadow than for Kerr spacetime, corresponding to a smaller (larger) value of $\delta_{s}$.  

Instead of the observables $R_s$ and $\delta _s$ introduced by Hioki and Maeda \cite{maeda}, other observables can be used from which the same information will be obtained, for example, those defined by Schee and Stuchlik \cite{schee}. We think that $R_s$ and $\delta _s$ have a more direct physical meaning, the first one giving an estimate of the size of the shadow and the other one about its deformation with respect to the non-rotating case, so we have adopted them for the present work.

\section{Naked singularity shadow}

\begin{figure}[t!]
\begin{center}
\includegraphics[width=0.4\linewidth]{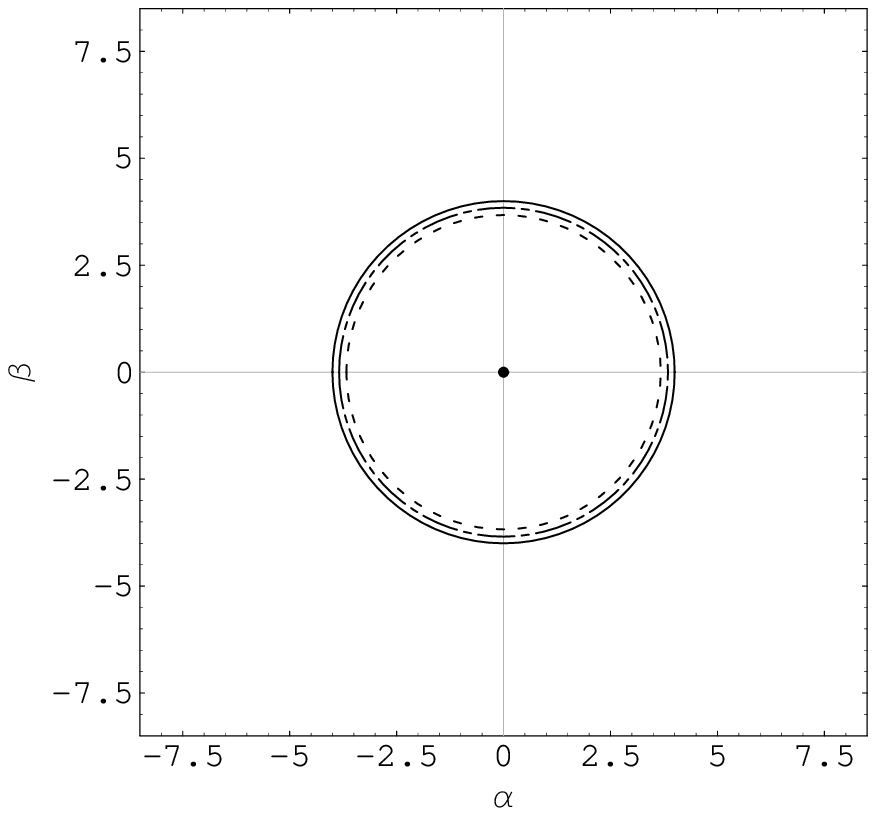}
\hspace{0.05\linewidth}
\includegraphics[width=0.4\linewidth]{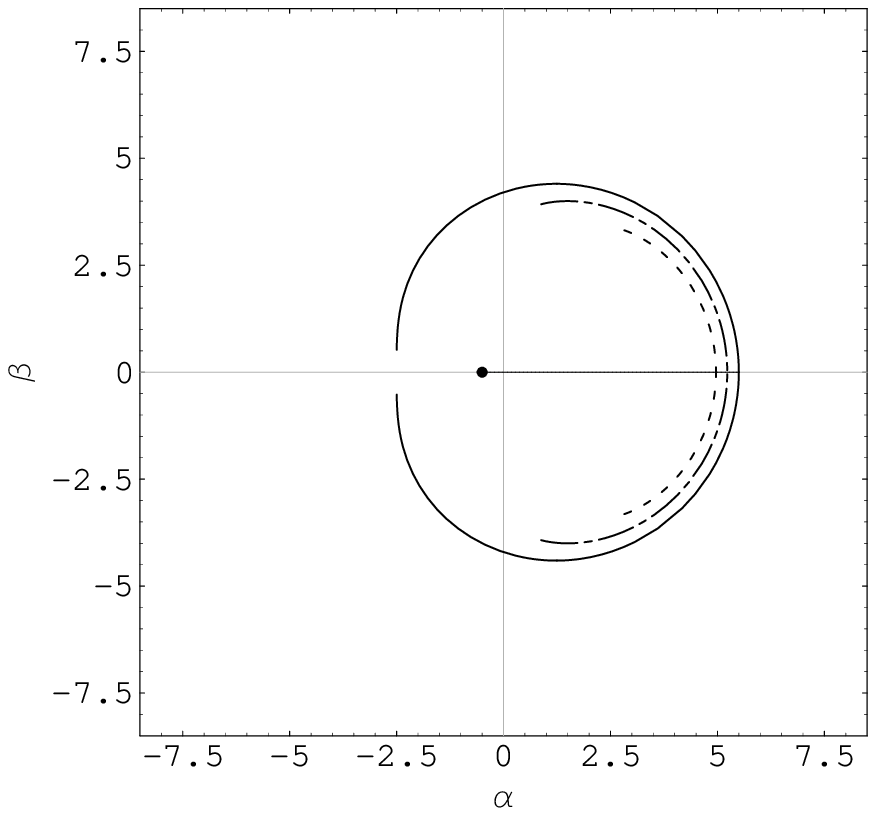}\\
\includegraphics[width=0.4\linewidth]{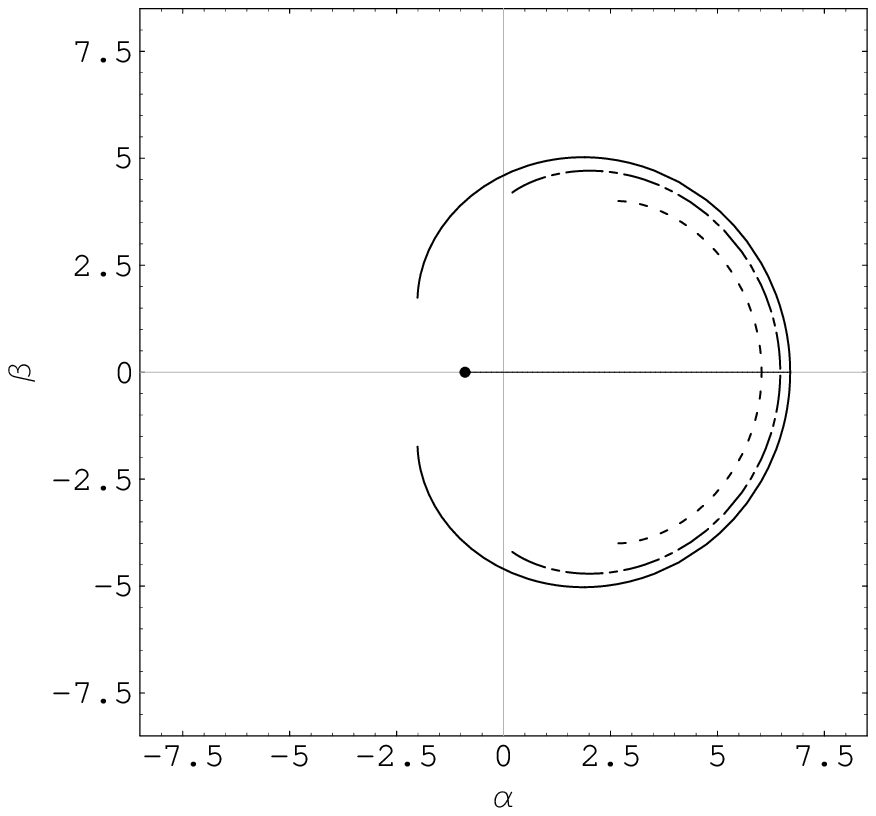}
\hspace{0.05\linewidth}
\includegraphics[width=0.4\linewidth]{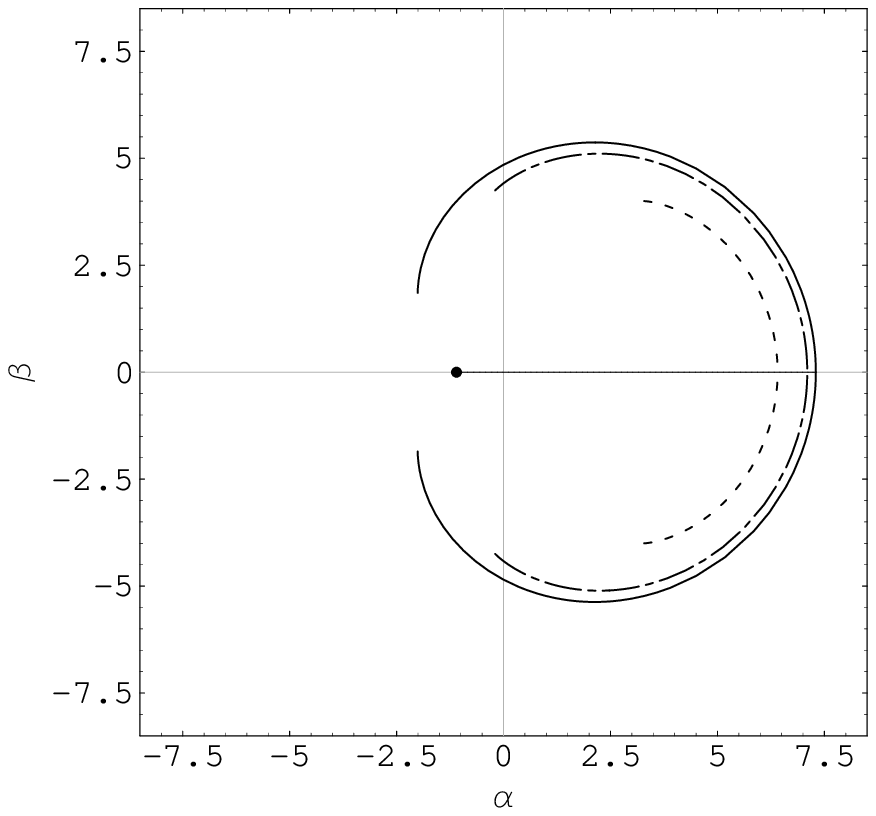}
\end{center}
\caption{Shadow cast by a naked singularity situated at the origin of coordinates with inclination angle $\theta _0=\pi /2$, having a rotation parameter $a$ and a tidal charge $Q$. Upper row, left: $a=0$, $Q= Q_{c} +10^{-5}= 1 + 10^{-5}$ (full line), $1.07$ (dashed-dotted line), and $1.125$ (dashed line). Upper row, right: $a=0.5$, $Q= Q_{c} + 10^{-5} =0.75 + 10^{-5}$ (full line), $1$ (dashed-dotted line), and $1.2$ (dashed line). Lower row, left:  $a=0.9$,   $Q= Q_{c} +10^{-5}= 0.19 + 10^{-5}$ (full line), $0.5$ (dashed-dotted line), and $1$ (dashed line). Lower row, right:  $a=1.1$,   $Q= Q_{c} +10^{-5}= -0.21 + 10^{-5}$ (full line), $0.1$ (dashed-dotted line), and $1$ (dashed line). All quantities were adimensionalized with the mass of the naked singularity (see text).}
\label{fig4}
\end{figure}

\begin{figure}[t!]
\begin{center}
\includegraphics[width=0.4\linewidth]{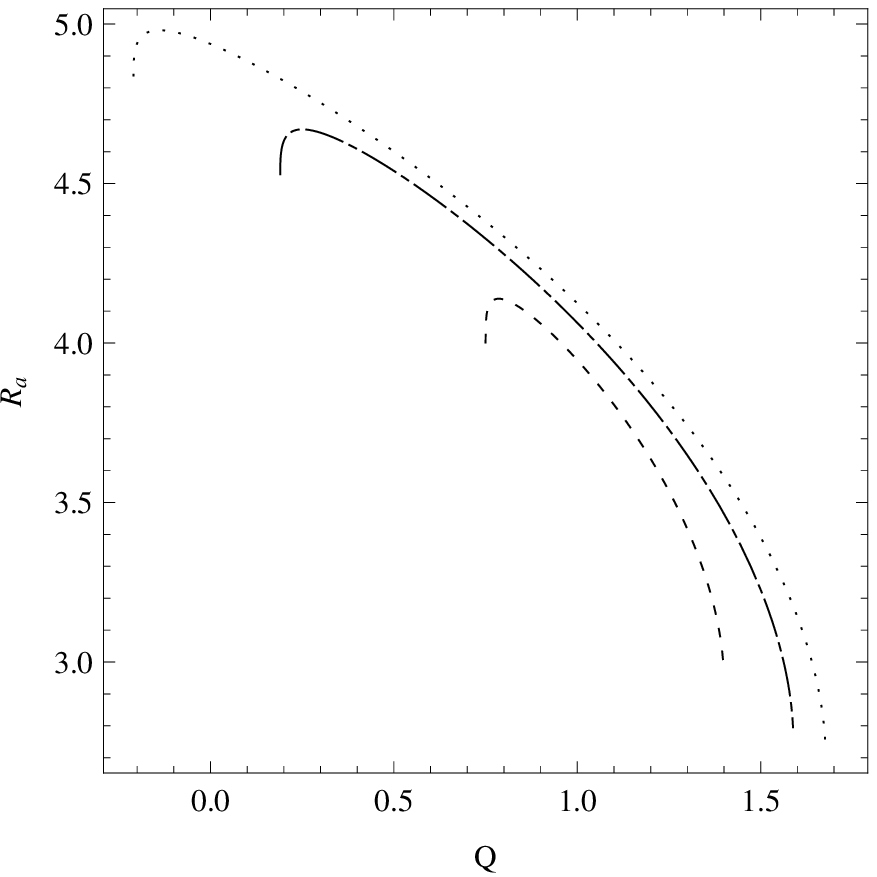}
\hspace{0.05\linewidth}
\includegraphics[width=0.4\linewidth]{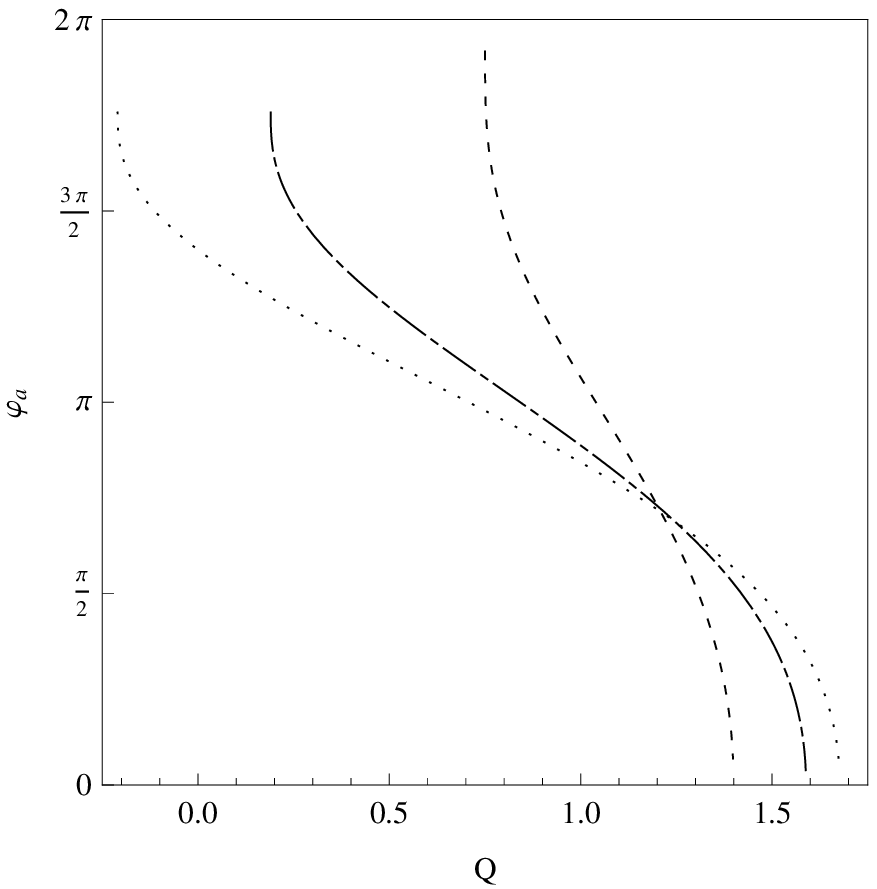}
\end{center}
\caption{Observables $R_{a}$ and $\varphi _{a}$ as functions of the tidal charge $Q $, corresponding to the shadow of a naked singularity situated at the origin of coordinates with inclination angle $\theta _0=\pi /2$, and spin parameters $a=0.5$ (dashed line), $a=0.9$ (dashed-dotted line), and $a=1.1$ (dotted line).}
\label{fig5}
\end{figure}

\begin{figure}[t!]
\begin{center}
\includegraphics[width=0.4\linewidth]{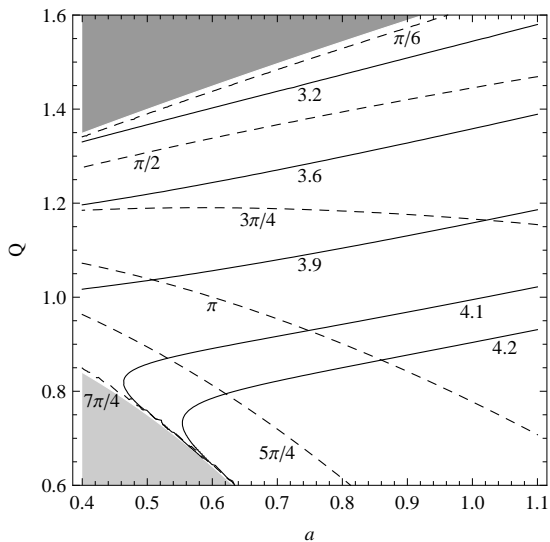}
\end{center}
\caption{Contour plots of the observables $R_{a}$ (full line) and $\varphi _{a}$ (dashed line) in the plane $(a,Q)$, for a naked singularity situated at the origin of coordinates with inclination angle $\theta _0=\pi /2$. Each curve is labeled with the corresponding value of $R_a$ or $\varphi _a$. In the light gray zone the singularity is covered by the horizon, while in the dark gray zone the arc is not present due to the absence of the photon sphere.}
\label{fig6}
\end{figure}

When $Q>Q_c=1-a^2$, i.e. if $a^2 +Q >1$, from Eq. (\ref{horizon}) it is easy to see that the horizons fade out; then for $a=0$ a point naked singularity appears at the origin, while for $a \neq 0$ a ring shaped naked singularity is obtained. The apparent shape changes radically with respect to the black hole case. When $a=0$, i.e. non-rotating naked singularity with a tidal charge, the zone inside the silhouette is not dark, and the shadow consists of a dark circumference with a dark point at the center, corresponding to the photon sphere and to the singularity, respectively. If $a\neq 0$, the behavior is similar to the naked singularity in Kerr or Kerr-Newman geometries, studied in Refs. \cite{devries,maeda}. The unstable spherical photon orbits with a positive radius give way to an open arc instead of a closed curve; the photons near both sides of the arc can reach the observer due to the nonexistence of the horizon. The finishing points of the arc have celestial coordinates $(\alpha , \beta )$ given by Eqs. (\ref{alphapsi1}) and (\ref{beta2}), where $\xi$ and $\eta $ are evaluated at $r_{min}=1-(1-a^2-Q)^{1/3}$, which is obtained in a similar way as in Ref. \cite{devries}. For naked singularities, it makes sense the extension of the geometry to $r<0$. The unstable spherical photon orbits with a negative radius construct a dark spot: the observer will never see the light rays from such directions because they escape into the other infinity by passing through the inside of a singular ring. The dark point by the principal null geodesics appears inside the spot. When the observer is on the equatorial plane ($\theta _0 =\pi /2$), the same arc exists but the dark spot disappears. The reason is that the light rays in the direction of negative radius will always hit on the ring singularity. Those null geodesics result in a straight line, with its endpoint corresponding to the principal null geodesics. If the inclination angle $\theta _0$ is varied from $0$ to $\pi /2$, the dark spot shrinks to a single point, which is the endpoint of the straight line (the principal null geodesics).

As pointed out above, the shadow of a rotating naked singularity consists of the arc and the dark spot or the straight line. The more interesting shape is the arc,  from which we can define two observables for the shadow: the radius $R_a$ and the central angle $\varphi _a$, as it was done in Ref. \cite{maeda} for Kerr geometry. The value of $R_a$ is defined as the radius of the circumference passing by the middle point of the arc and by the two points where the arc finishes. The observable $\varphi _a$ is determined by the angle subtended by the arc, seen from the center of the circumference used to define $R_a$.  As in the black hole case, when the inclination angle is known, if $R_a$ and $\varphi _a$ are obtained from observations, the associated contour curves of constant $R_a$ and $\varphi _a$ in the plane $(a,Q)$ can be used to obtain the corresponding values of the rotation parameter $a$ and the tidal charge $Q$.

For the reasons stated in the previous section, the more interesting case to analyze is the one corresponding to an observer in the equatorial plane. In Fig. \ref{fig4}, we show the shadows of naked singularities situated at the origin of coordinates with inclination angle $\theta _0=\pi /2$, for rotation parameters $a=0$ (upper row, left), $a=0.5$ (upper row, right), $a=0.9$ (lower row, left), and $a=1.1$ (lower row, right), for several values of the tidal charge $Q$, larger than the corresponding critical value $Q_{c}$. When $a=0$ we see that the shadow is a circumference with a radius that decreases with the tidal charge and a central dark point. If $a\neq 0$, we have that the shadow consists of a dark arc plus a line which finishes in the dark point in the figures. The arc closes when the tidal charge is near to $Q_c$ and opens up as the charge increases. 

In Fig. \ref{fig5}, the observable $R_{a}$ is shown as a function of the tidal charge $Q$, for several values of the rotation parameter of the naked singularity: $a=0.5$ (dashed line), $a=0.9$ (dashed-dotted line), and $a=1.1$ (dotted line). The behavior of $R_{a}$ is similar for the three values of $a$, having a small growth for values of $Q$ near $Q_{c}$, and then a decrease. The initial growth is related to the particular shape of the shadow generated by the rotation, and the definition of the circumference of reference. On the other hand, the decrease continues until the arc generated by the photons belonging to the photon sphere turns to a single point and then disappears indicating the loss of the photon sphere. As expected, the radius is also an increasing function of the rotation parameter $a$ which can be deduced from the position of the curves for different values of $a$ in the figure. The angle subtended by the end points of the shadow, measured from the center of the circumference of reference is a decreasing function of $Q$. In Fig. \ref{fig5}, the plot of the angle $\varphi_{a}$ as a function of $Q$ can be seen for $a=0.5$, $a=0.9$ and $a=1.1$. The  limiting values are $\varphi_{a}=2 \pi$ and $\varphi_{a}=0$, corresponding to the critical value of the charge for which the singularity is covered by the horizon, and to the disappearance of the photon sphere, respectively. For $Q \lesssim 1.25$ the angle is a decreasing function of the rotation parameter, while for $Q \gtrsim 1.25$ the behavior is the opposite. The values of the rotation parameter $a$ and the tidal charge $Q$ can be obtained from the intersection point of the curves with constant $R_a$ and $\varphi _a$ in the plane $(a,Q)$; to exemplify this, some representative contour curves are displayed in Fig. \ref{fig6}.

\section{Discussion}

In this article, we have extended previous studies \cite{aliev2,schee} of gravitational effects corresponding to rotating black holes in Randall-Sundrum braneworld cosmology. In particular, we have analyzed how the shadow of the black hole is distorted by the presence of the tidal charge. From the observable $R_s$, we have found that the size of the shadow for a fixed value of the rotation parameter $a$ decreases with the tidal charge $Q$, resulting in a larger shadow than in Kerr geometry for negative $Q$ and a smaller shadow for positive $Q$. The deformation of the shadow, characterized by the observable $\delta _s$, increases with $Q$, so a less distorted silhouette is obtained for negative $Q$ and a more deformed one for positive $Q$.

The angular size of the shadow can be estimated by using the observable $R_{s}$ to obtain the angular radius $\theta _s = R_{s}M/D_{o}$, with $D_{o}$ the distance from the observer to the black hole. It is easy to see that $\theta _s = 9.87098\times 10^{-6} R_{s}(M/M_{\odot}) ( 1 \, \text{kpc}/D_{o})$ $\mu \mathrm{as}$. The observable $\delta _s$ (\%) gives an idea of how the silhouette is deformed with respect to the non-rotating case. For the supermassive black hole Sgr A* at the Galactic center, we have $M=4.3 \times 10^{6}M_{\odot}$ and $D_{o}=8.3$ kpc \cite{guillessen}; then,  under the assumption that the observer is in the equatorial plane, we obtain 
\begin{center}
\begin{tabular}{|c|c|c|c|c|c|c|c|c|}
\hline
$a$ & \multicolumn{4}{|c|}{$0$}  & \multicolumn{4}{|c|}{$0.9$}\\
\hline
$Q$ & $-0.5$ & $-0.1$ & $0$ & $0.1$ & $-0.5$ & $-0.1$ & $0$ & $0.1$\\
\hline
$\theta _s (\mu \mathrm{as})$ & $28.605$ & $27.006$ & $26.572$ & $26.120$ & $28.612$ & $27.018$ & $26.586$ & $26.136$ \\
\hline
$\delta _s (\%) $  & $ 0$ & $0$ & $ 0$ & $0$ & $7.45$ & $11.8$ & $13.9$ & $17.2$\\
\hline
\end{tabular}
\end{center}
where the values of $a$ and $Q$ were chosen only for illustrative purposes. We see from the table that resolutions of less than $1 \, \mu \mathrm{as}$ are needed in order to extract useful information from future observations of the shadow of the Galactic supermassive black hole.

We have also studied the shadows of naked singularities, corresponding to the condition $a^2 + Q >1$, for which the horizons associated to the braneworld metric considered in this paper fade out. For a rotating naked singularity, in the case that this kind of object exists in nature, the shadow has two parts: the arc and the dark spot or the straight line. The more interesting shape is the arc, which may not be observable because it is one-dimensional. In a realistic scenario, however, the neighborhood of the arc will also be darkened to be observed as a dark \textit{lunate} (i.e. crescent moon shaped) shadow, as it was pointed out in Ref. \cite{maeda} for the Kerr metric. Then, we would have a chance to observe the shadow of a naked singularity, but it will be even more difficult than in the case of black holes. For a given value of the rotation parameter $a$, the size of the arc is slightly smaller than the size of the distorted dark disk corresponding to a tidal charge satisfying $a^2 + Q \le 1$, for which the horizon is present.

The observation of black hole (or naked singularity) apparent shapes is a major goal in observational astrophysics, since those shadows correspond to a full description of the near horizon region, without any theoretical assumption concerning the underlying theory or astrophysical processes in the black hole surroundings. In the near future, some observational facilities, most of them space-based, will be fully operational and will be able to measure in the radio and X bands. Two of them are worth mentioning, namely, RADIOASTRON and MAXIM. The first one is a space-based radio telescope launched in July 2011. It will be capable of carrying out measurements with high angular resolution, about $1-10 \, \mu \mathrm{as}$
\cite{zakharov,webradio}. On the other hand, the MAXIM project is a space-based X-ray interferometer with an expected angular resolution of about $0.1 \, \mu \mathrm{as}$ (see \cite{webmaxim} for further details). These instruments will be able to resolve the shadow of the supermassive Galactic black hole, which together with other observations, would serve to obtain its parameters in the near future. It seems that more subtle effects, like the comparison of different models of black holes corresponding to alternative theories, will require a second generation of instruments to be available in the future. If the braneworld model provides a suitable description of cosmology, one would expect a small (and negative, as stated above) value of the tidal charge, so a very large resolution will be necessary for the observation of the effects discussed in this work.

\begin{acknowledgments}

This work was supported by CONICET and Universidad de Buenos Aires. We want to thank an anonymous referee for careful reading of the manuscript and for helpful comments.

\end{acknowledgments}

\end{document}